\documentclass[11pt]{article}
\usepackage{epsfig}
\usepackage{geometry}                
\usepackage{graphicx}
\usepackage{amssymb}
\usepackage{epstopdf}
\title{Spectral analysis of the background in ground-based, 
   long-slit spectroscopy}
\author{Fr\'ed\'eric Zagury \thanks{e-mail:fzagury@wanadoo.fr}
\\ Institut Louis de Broglie, 23 rue Marsoulan, 75012 Paris, France}

\begin{document}
\maketitle

\begin{abstract}
This paper examines the variations, because of atmospheric extinction,
of broad-band visible spectra, obtained from long-slit spectroscopy, 
in the vicinity of some stars, nebulae, and one faint galaxy.
\end{abstract}
    \section{Introduction}
In a preceding paper (Zagury \& Goutail 2003) we have analyzed 
the extinction effects of the atmosphere on 
sunlight, as they can be observed by a balloon-born spectrometer.
The study specified what should be the radiation field, due to 
the sun, in any point of a clear atmosphere, 
and we were able to derive its analytical form.

The effect of the atmosphere 
on the visible continuum of an external source 
of light is three-fold:
Rayleigh extinction by the gas (nitrogen mainly, $\propto 
e^{-g/\lambda^{4}¥}$); extinction by aerosols ($\propto 
e^{-a/\lambda}$); and the broad-band ozone absorption (Chappuis 
bands).

In the complete forward direction, direct light from the source 
will be extinguished according to these three effects.
In near-forward directions, direct light from the 
source is still present due to refraction and turbulence in the atmosphere, 
but, as we move away from the direction of the source,
drops in intensity, and the source's 
light scattered by aerosols (with a spectrum $\propto 
e^{-a/\lambda}/\lambda$) will become discernible.
Farther away, when direct light from the source 
is negligible, forward scattering by aerosols and 
ozone absorption will be the only remaining processes at work.
At very large angles, forward scattering is no more effective; 
isotropic Rayleigh scattering by the gas ($\propto e^{-g/\lambda^{4}¥}/\lambda^{4}¥$) 
and the ozone absorption feature should dominate the spectrum.
A first application of this work was presented by Zagury \& Fujii 
(2003) who were 
able to fit spectra of a red horizon at sunrise.

In this paper I shall be concerned with the continuum of spectra 
observed in the vicinity of different types of objects: stars, nebulae 
and galaxies.
The spectra, obtained through long-slit spectroscopy, are considered 
after correction for the instrumental response, but before absolute 
calibration (no correction of atmospheric extinction), so that 
the effects mentioned above can be recognized and eventually separated.

Sections~\ref{data}  and \ref{pres} present the different observations used in 
the paper and a series of notes collected during the investigations 
for 
its preparation.
Observations are analyzed in Sections~\ref{fdstar} and \ref{fdneb} and discussed in 
Section~\ref{dis}.
   \section{Observations} 
   \label{data}
The observations used in this paper all emanate from long slit 
spectroscopy.
Some were observed with the FAST spectrograph (Fabricant \emph{et al.} 1998) 
using the 1.5~m Tillinghast telescope of the Fred Lawrence Whipple Observatory
(Mount Hopkins, U.S.A).
The others come from the Kwitter-Henry database (http://cf.williams.edu/public/nebulae/browse.cfm; 
Kwitter \& Henry 2002), and were observed with the 2.1~m telescope at Kitt Peak National 
Observatory (KPNO).
I will also refer to night sky spectra observed by Phil Massey and 
Craig B. Foltz (Massey \& Foltz 2000), also obtained at Kitt Peak and Mount 
Hopkins.

Table~\ref{tbl:pos} gives the positions and the visual magnitudes of 
the objects (central stars for nebulae) used in the paper.
Spectrum `$x$' of a 2-D array will be noted $sx$.
Vertical dotted lines on the plots give the  limits of
the ozone absorption region.
\subsection{Mount Hopkins FAST observations} \label{fast}
The reference paper for the FAST spectrograph is Fabricant \emph{et al.} 
(1998).
The configuration used for the observations is a 300 line mm$^{-1}$ 
grating.
The slit samples a region $\sim 3'$ long and 3'' wide on the sky.
The wavelength coverage goes from  $3660\,\rm \AA$ to $7530\,\rm \AA$, 
with a resolution of $1.47\,\rm\AA/pixel$.
The slit is divided into 320 pixels, with a spatial resolution of $0.61$''/pixel.
Note that the level of undesired scattered light (reflections from the refractive optics and 
the CCD, imperfections in the reflective surfaces and scattering of 
light outside first order from the spectrograph case) is estimated to 
be less than $1\% $ of the spectra (see the FAST specifications at 
http://tdc-www.harvard.edu/instruments/fast/).

The observations (Table~\ref{tbl:fast}) of the three stars (HD23302 
($17\,\tau$), HD44113, HD44179) and of the Red Rectangle 
nebula are extracted from a larger set 
observed by Dr. Lucas Macri during three observing runs, in 2001, 
December 21 and 22, 
2002, February 9, and 2003, March 26.
Observing conditions during the three runs were good, with a seeing 
in the range 1-2''. 
Galaxy UGC11917 was observed by Dr. I. Ginsburg on 
2002, October 31.

The data reduction was done by Dr. Susan Tokarz (Tokarz \& Roll 1997).
Wavelength calibration is made with a HeNeAr lamp.
A final 2-D array is extracted for each observation, 
which is dark-corrected and corrected for the 
instrumental response, but not for atmospheric extinction.
The 2-D arrays have been binned by two or by four during the process, along the spatial 
dimension, yielding 
a final spatial resolution of $1.2$''/pixel for the binned by two 
spectra and $2.4$''/pixel for the binned by 4.
All 2-D FAST spectra, except those from February 2002, are binned by 
two.
1-D spectra of point source objects are usually extracted from the 2-D 
arrays (Section~\ref{1d}). 
\subsection{Kitt Peak observations } \label{kp}
The Kwitter-Henry database (http://oit.williams.edu/nebulae/browse.cfm) 
proposes a collection of 88 
nebular spectra, in the [$3600\,\rm\AA$, $9600\,\rm\AA$] wavelength 
range.
Dr. Kwitter gave me access to the 2-D arrays, at the same early stage of data 
reduction as for the Mount Hopkins observations (i.e. corrected for 
the instrumental response but not for atmospheric extinction), of 
three planetary nebulae: NGC2022, NGC6309, and NGC6891.

These spectra, as for the preceding FAST spectra, are issued from 
long-slit spectroscopy (see Kwitter \& Henry 2001).
The slit is 5'' wide, 285'' long in the E-W direction.
The spatial resolution is 0.78''/pixel.
The spectral resolution is $1.49\,\rm \AA /pixel$.
Only the blue spectra will be used, which 
range from $3550\,\rm \AA$ to $6800\,\rm \AA$ for the NGC2022 observation, 
from $3640\,\rm \AA$ to $6780\,\rm \AA$ for NGC6309 and NGC6891 
observations. 
Contrary to the FAST spectra, these spectra are unbinned.

The observations were obtained at KPNO with the $2.1\,$m 
telescope, 1999 July 1 for NGC6309 and NGC6891 (Kwitter \& Henry 2001), 1996 
December 8 for NGC2022 (Kwitter \emph{et al.} 2003) (Table~\ref{tbl:kittpeak}).
\section{Data overview} \label{pres}
\subsection{2-D spectra of stars} \label{st2d}
\begin{figure*}[p]
\resizebox{1.\textwidth}{!}{\includegraphics{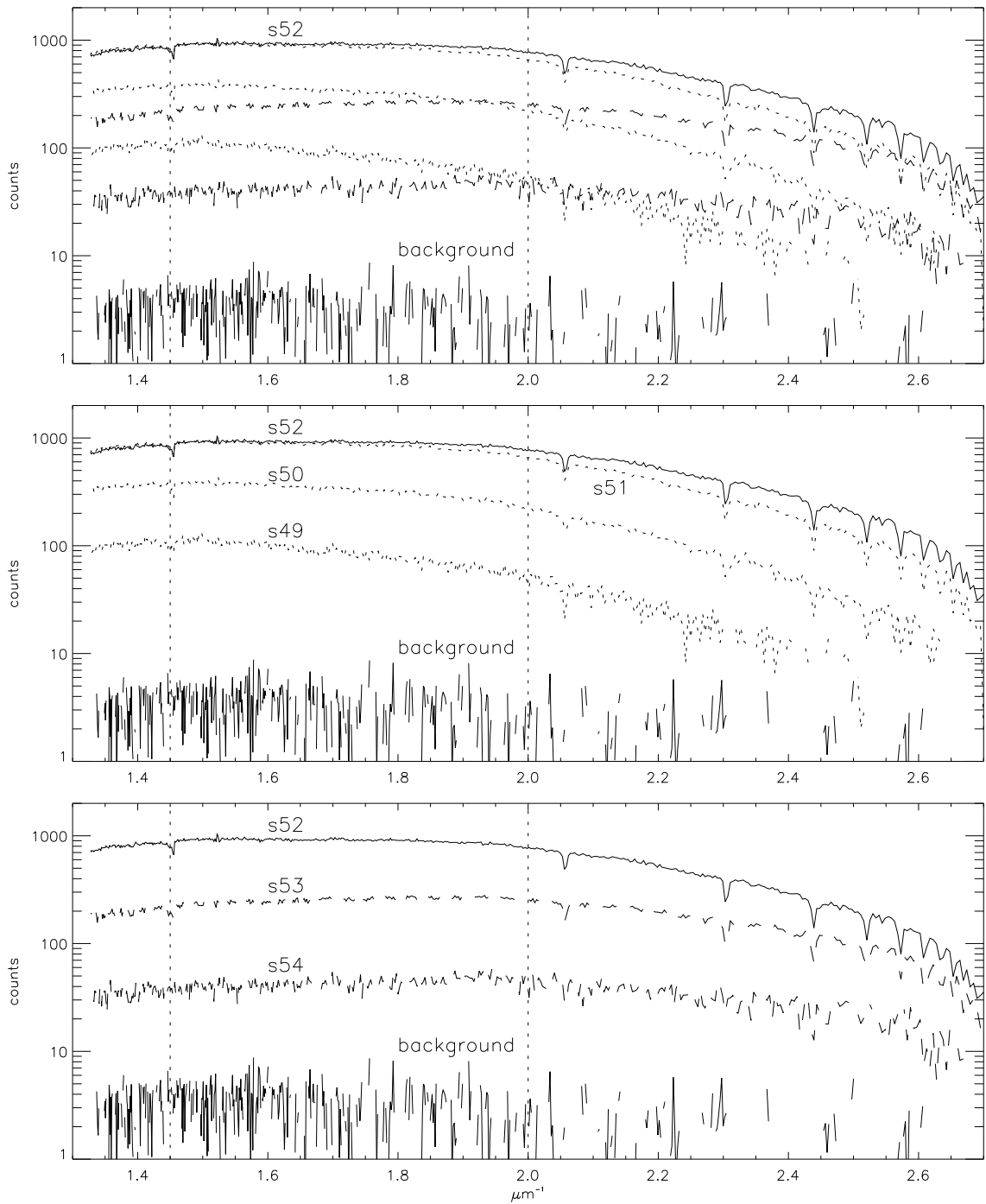}} 
\caption{HD44179 (Red Rectangle), first March 2003 observation.
The pixel with maximum signal (main pixel of the 2-D array) is pixel 52.
Most of the direct starlight is concentrated in pixels 49 to 54.
The spectra then reach a common shape (the background).
} 
\label{fig:rr2d}
\end{figure*}

Figure~\ref{fig:rr2d} plots spectra from the 2-D array, after instrumental 
calibration and before correction for atmospheric extinction, of one of 
the stars (HD44179, first March 2003 observation).
Although the direct light from HD44179 is mainly concentrated in 
one or two pixel(s) (pixels 52 and 51 in this example), 
which corresponds to the seeing, and fixes the direction of the star,
it is distributed over several pixels (here pixels 49 to 54, 
which represent $\sim 7$" on the sky),
with a decrease in intensity of a factor of 
100 between pixel 52 and pixels 49 or 54.

The main cause of the spread of the signal over several central pixels 
is the dispersion of light in the atmosphere 
due to turbulence and refraction.
The effect of atmospheric extinction on these spectra is manifest 
in the blue, 
the long wavenumber (short wavelength), $1/\lambda>2\rm\mu m^{-1}$, 
bending of the spectra being due 
to Rayleigh extinction by nitrogen ($\propto e^{-g/\lambda^{4}¥}$).

\begin{figure*}[]
\resizebox{1.\textwidth}{!}{\includegraphics{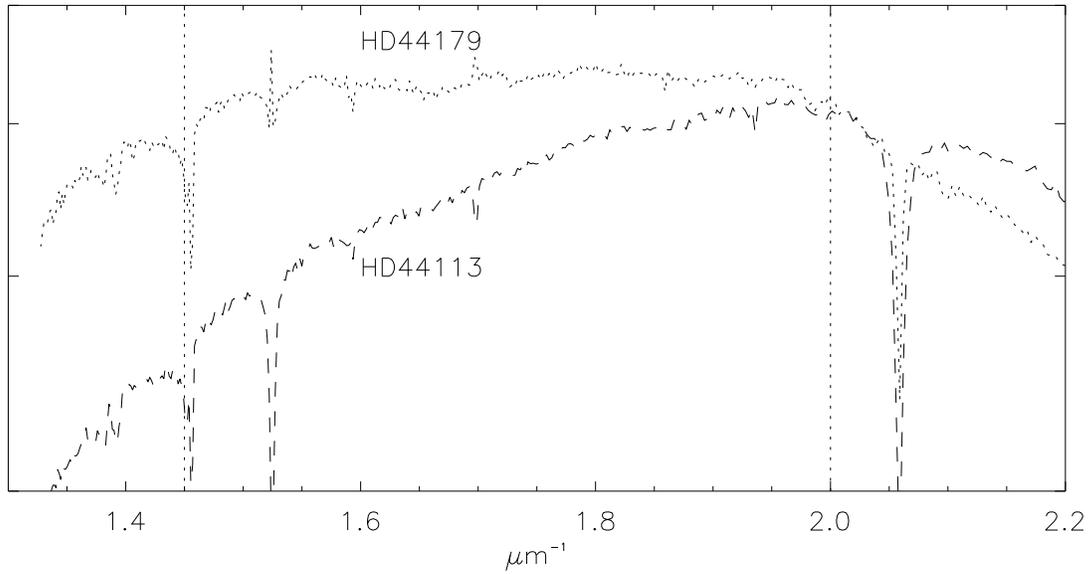}} 
\caption{1-D spectra of February 2002 observations of HD44179 and 
HD44113 (arbitrary scaling).
Ozone absorption is responsible for the flattening of HD44179 spectrum and the
depression between $\sim 1.5\,\mu\rm m^{-1}$ and $\sim 1.9\,\mu\rm m^{-1}$.
Although present in the same proportion, it is  more difficult to detect in the observation of HD44113
because of the steep rise of the spectrum.
Rayleigh $1/\lambda^4$ extinction is responsible of the 
decrease longward of $2\,\rm\mu m^{-1}$.
} 
\label{fig:oz}
\end{figure*}

Ozone absorption affects the $[1.5\rm\mu m^{-1}¥,\, 2\rm\mu m^{-1}]$ wavenumber 
region (Figure~\ref{fig:oz}), but is not always easy to distinguish.
Modifications of the sun spectrum in Zagury \& Goutail (2003), due to ozone, 
are perceptible for column densities 
of $\sim 9\,10^{18}$~molecule/cm$^{2}$, and are important when the column 
density reaches a few $10^{19}$~molecule/cm$^{2}$.
Zenithal ozone column densities observed from earth are of order 
300~Dobson (1~Dobson$=2.69\,10^{16}$~molecule/cm$^{2}$). 
For zenithal distances of order 
$\theta\sim 45^{\circ}$, as for HD44113 and HD44179 observations,
the ozone column density will be 
$\sim 8\,10^{18}/\cos\theta = 10^{19}$~molecule/cm$^{2}$.
Hence, our observations are in the transition region where ozone 
absorption starts to be significant. 

The differences between the main spectra of an observation (pixels 49 
to 55 in Figure~\ref{fig:rr2d}) can be appreciable, 
and there is an order in the successive modifications of the spectrum's shape.
Moving from spectrum $s49$ to $s54$, there is a progressive decrease (in 
absolute value) of 
the average slope of the spectra; this
coincides with, 
and may be due to, the orientation of the slit in the azimuth-elevation 
plane (pixel numbers increase with pixel altitudes).
This is observed in all star observations.

\begin{figure*}[]
\resizebox{1.\textwidth}{!}{\includegraphics{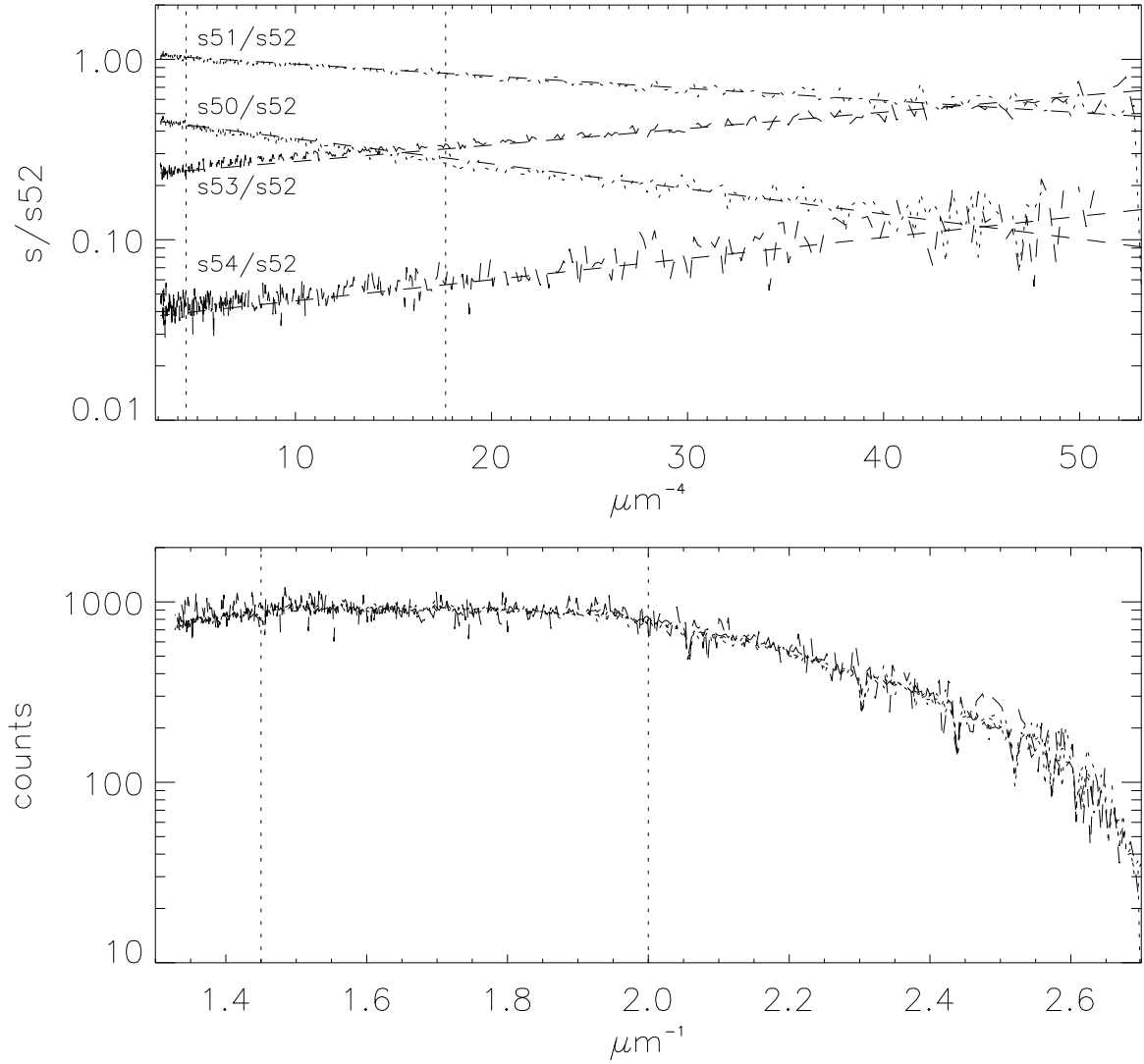}} 
\caption{Relations between the spectra in the second March 2003 observation
of HD44179.
\emph{Top:} ratios of $s51$, $s50$, $s53$, $s54$ to $s52$ are an exponential of $1/\lambda^{4}$¥.
\emph{Bottom:} $s52$ is superimposed by 
$0.9e^{0.015\lambda^{-4}¥}¥s51$,
$2e^{0.032\lambda^{-4}¥}e^{10^{19}\sigma¥}¥s50$,
$4.5e^{-0.021\lambda^{-4}¥}e^{-2\,10^{19}\sigma¥}¥s53$,
and $28.6e^{-0.027\lambda^{-4}¥}e^{-2\,10^{19}\sigma¥}¥s54$.
} 
\label{fig:2drel}
\end{figure*}

The relationship between these spectra often takes a simple 
analytical expression.
Main spectra in Figure~\ref{fig:rr2d} for instance relate to each 
other by an exponential of $1/\lambda^{4}$ (as for the second March 
2003 observation of HD44179, Figure~\ref{fig:2drel}) 
-which could be due to differences in extinction by the gas (though 
the coefficients of the exponent are high compared to what was found 
in Zagury \& Goutail 2003)-
and a small correction for absorption by ozone.
The transformations hold good for other 
observations made during the same night and in the same region.

The angular extent over which direct light from a star
can be detected depends on the observation.
Its diameter, centered on the direction of the star, 
is of $\sim 12$" and $\sim 14$" for the HD23302 and HD44179 
December 2001 observations.
For the March 2003 observation, the extent keeps the same order of 
magnitude for 
HD23302, but, as mentionned above, it is divided by 2 for HD44179.
February observation of HD44179 gives, as for March 2003, $\sim 7$".
These important differences, for HD44179, between December 2001 and  
the other runs, could be related to the exposure 
time (February 2001 and March 2003 exposures are one third of December 
2001's), maybe an effect of the time-dependence of turbulence? 

Contamination of the main spectra ($s49$ to $s54$) by starlight scattered in the 
atmosphere should be negligible since the scattered light 
should not be more than one per cent of the star's flux 
(Zagury \& Goutail 2003; Zagury 2001).

Moving away from the central pixel of a 2-D array, the signal quickly decreases to 
a spectrum whose  shape remains constant (the background, Figure~\ref{fig:rr2d}), and 
is the same on both sides of the central pixel. 
The background spectrum of Figure~\ref{fig:rr2d} is the average 
of eight spectra ($s43$ to $s46$, $s57$ to $s60$),
taken at a mean distance of $\sim 9''$ from HD44179.

\begin{figure*}[]
\resizebox{1.\textwidth}{!}{\includegraphics{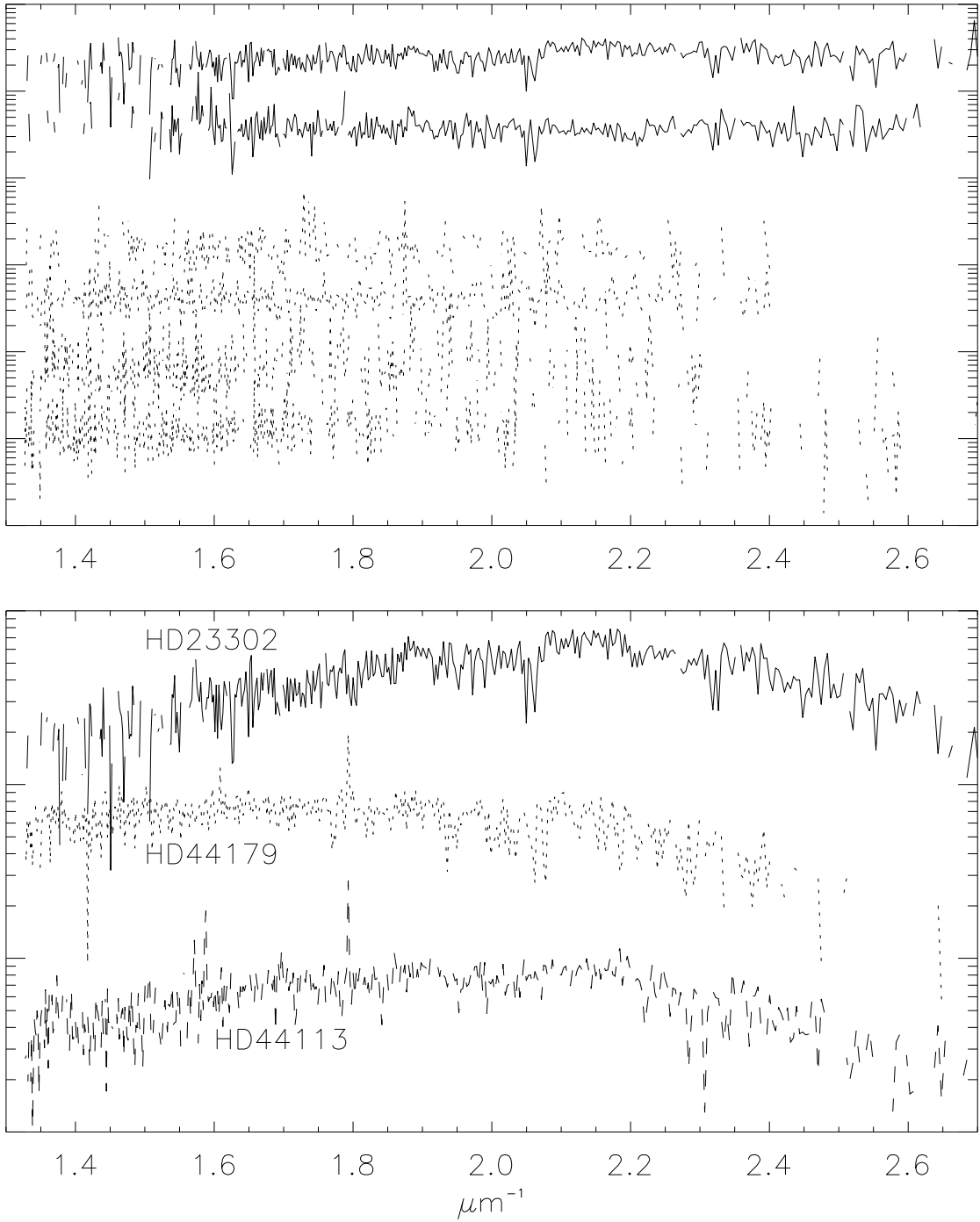}} 
\caption{
\emph{Top:} Ratios of backgrounds for different observations of HD44179 
(dots, ratios of all observations to the first December one) and HD23302 
(solid lines). The scaling is arbitrary.
\emph{Bottom:} Backgrounds  (on an arbitrary scale) of 
HD44113 (dashes), HD44179 (dots), and HD23302 (solid line), 
observations.
} 
\label{fig:fet}
\end{figure*}

Backgrounds of the same star, from different observations
(top plot of Figure~\ref{fig:fet}), are similar 
and differ at most by a slight change of 
slope (attributed to a difference in atmospheric 
optical depth).
The study of the backgrounds will thus be limited to one per star.
The background depends on which star is observed 
(Figure~\ref{fig:fet}, bottom plot).

The background slowly decreases with distance from the star so that its 
precise determination (and an eventual change of shape) 
is difficult at the edges of the slit.
The relationship which exists between the background, the atmosphere, 
and starlight, is investigated in Section~\ref{fdstar}. 
\subsection{1-D spectra} \label{1d}
A 1-D spectrum, before its absolute calibration, of a point-like object, 
is determined by adding the spectra of the pixels with larger signal [for 
the March 2003 observation of HD44179 (Figure~\ref{fig:rr2d}), $s49$ to 
$s55$ were added].

\begin{figure*}[p]
\resizebox{1.\textwidth}{!}{\includegraphics{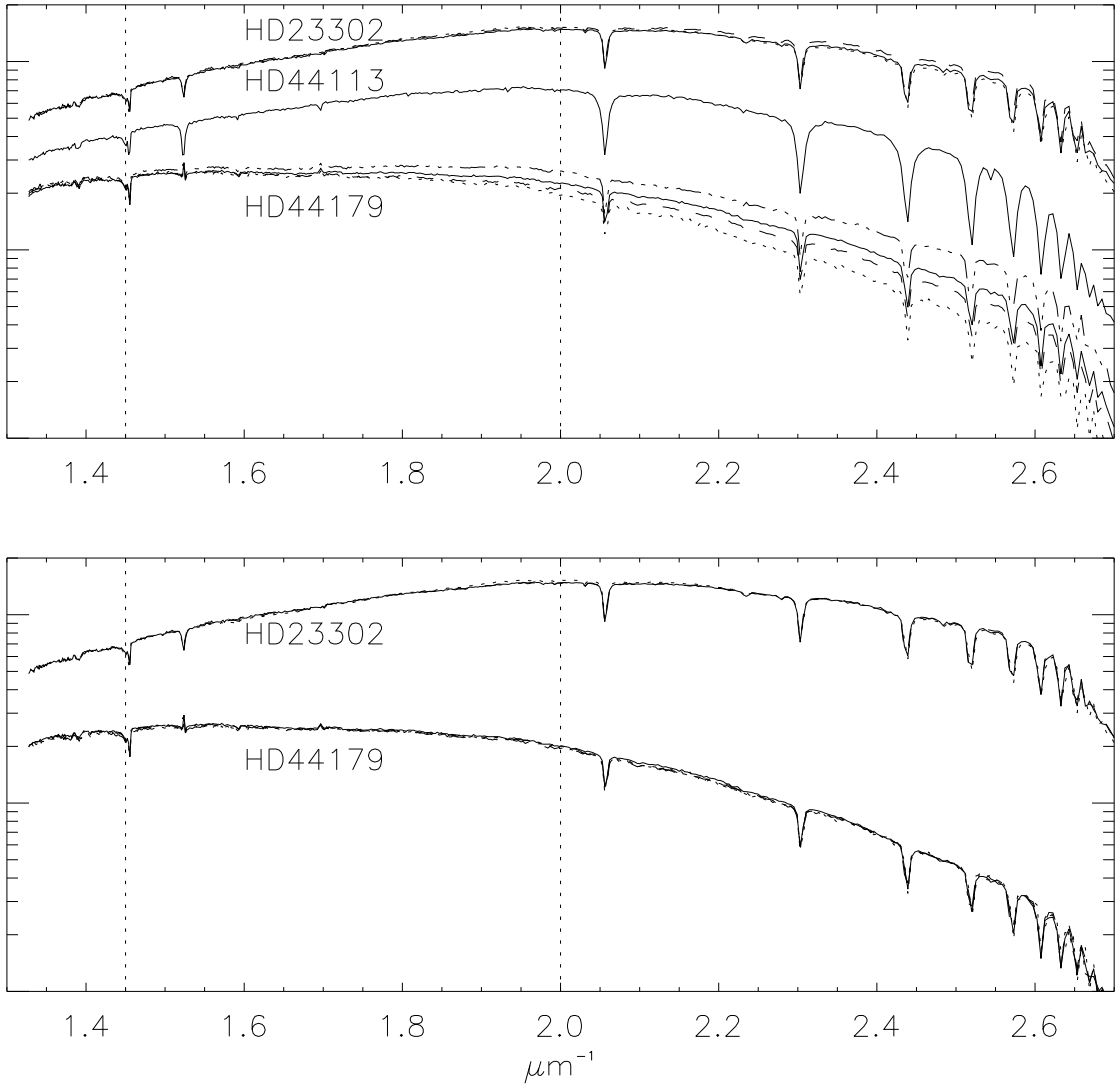}} 
\caption{\emph{top:} 1-D spectra, before atmospheric correction, of 
HD44179, HD44113, and of HD23302 observations.
Arbitrary scaling.
\emph{bottom:} 1-D spectra of HD44179 (resp. HD23302), from all 
observations, superimpose 
well after a correction for 
atmospheric extinction has been applied.
} 
\label{fig:etoiles}
\end{figure*}

Top panel of Figure~\ref{fig:etoiles} is a plot of the 1-D spectra, 
before their final calibration, 
of all the star observations used in this paper.

1-D spectra of the same object observed at different dates 
(top panel of Figure~\ref{fig:etoiles})
are related to each other by a transformation involving Rayleigh 
extinction and ozone absorption.
On the bottom plot of Figure~\ref{fig:etoiles}, 
second December 2001 and first March 2003 observations of HD23302 
reproduce  perfectly well the first December 2001 observation of the 
star, after
multiplication by $e^{-0.015\lambda^{-4}¥}$ and 
$e^{-0.0006\lambda^{-4}¥}e^{-0.6\,10^{19}\sigma¥}$
($\sigma$ is the ozone cross-section, expressed in cm$^{2}$).
The corrections to apply to the second December 2001, February 
2002,  and first March 2003 observations of HD44179, to match the first 
December 2001 spectrum, are: 
$e^{0.005\lambda^{-4}¥}e^{0.5\,10^{19}\sigma¥}$,
$e^{-0.007\lambda^{-4}¥}e^{0.5\,10^{19}\sigma¥}$,
$e^{0.012\lambda^{-4}¥}$.
\subsection{Extended objects} \label{nebspec}
Extended objects are four nebulae and one galaxy.
Their size is in the range 15''-25'', 
much smaller than the angular size the slit samples 
on the sky, but enough to encompass several pixels of the slit.

\begin{figure*}[p]
\resizebox{1.\textwidth}{!}{\includegraphics{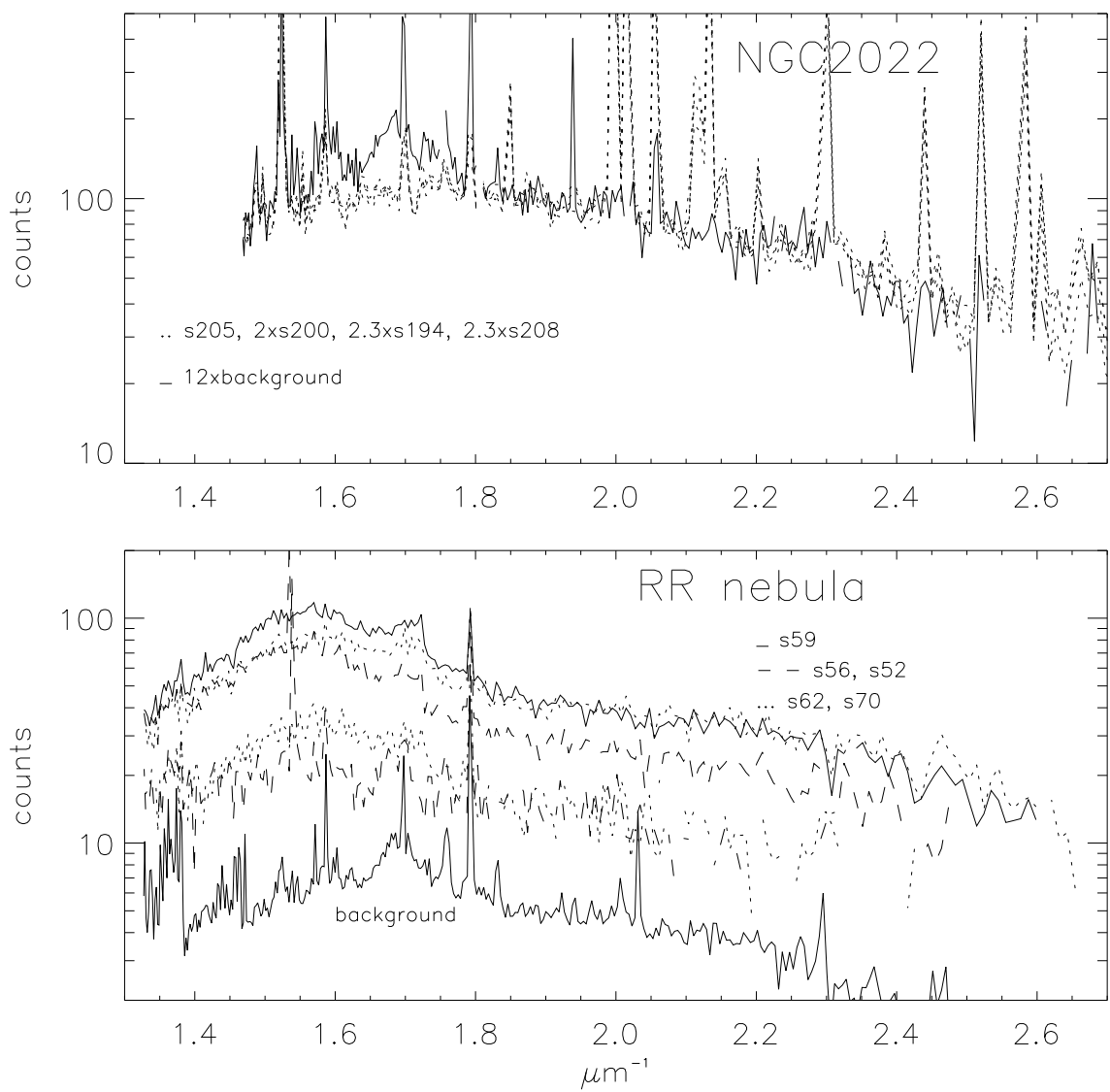}} 
\caption{Spectra from the 2-D arrays of NGC2022 and of the Red 
Rectangle nebula.
} 
\label{fig:neb2}
\end{figure*}

The slit for the Red Rectangle nebula was positionned $\sim 11.5$'' north 
of HD44179.
The spectra far above the background have the well known bump shape 
in their near infrared part (Figure~\ref{fig:neb2}, bottom plot).
The spectrum with maximum signal is $s59$.
Nearest spectra to $s59$ decrease short or long wavelengths first, 
according to which side of $s48$ the spectrum is, as in the observations of 
stars (Section~\ref{st2d}).
On a larger scale, the decrease of the spectra, from $s59$ to the background, is smooth, 
and does not show the strong variations found in the observation of 
stars.
The shape of the spectra differs from that of the background, 
at least in its red part, 
from $s51$ to $s72$, which represents $\sim 26''$ on the sky, in 
agreement with the spatial extent of the nebula (26-27''), 11'' north of 
HD44179.
The background, taken at the edges of the slit, is 
clearly different from the background found in the vicinity of HD44179.

NGC2022 was in principle observed $6.5$" south of the star (Karen 
Kwitter, private communication).
The header of the 2-D spectrum fits file, however, which normally indicates the position of the 
slit, gives an offset of 16'' (this would put the slit
on the far-edge, eventually out of the nebula).
All spectra from NGC2022 2-D array are proportional 
longward of $1.8\,\rm \mu m^{-1}$, with a progressive decrease on each 
side of the main pixel (pixel 205, Figure~\ref{fig:neb2}).
Moving away from $s205$, a bump-like feature 
progressively appears upon the spectra, along with the decrease of the 
spectrum, between $1.5\,\rm \mu m^{-1}$ 
and $1.8\,\rm \mu m^{-1}$.
In none of the spectra, on the possible extent (at most 19'') of the nebula around 
the direction given by pixel 205, is it possible to see  any specific 
feature, in the continuum, that could be attributed to the nebula.
It follows that the spectra, to the precision of these 
observations, contain no signal from the nebula.

\begin{figure*}[]
\resizebox{\textwidth}{!}{\includegraphics{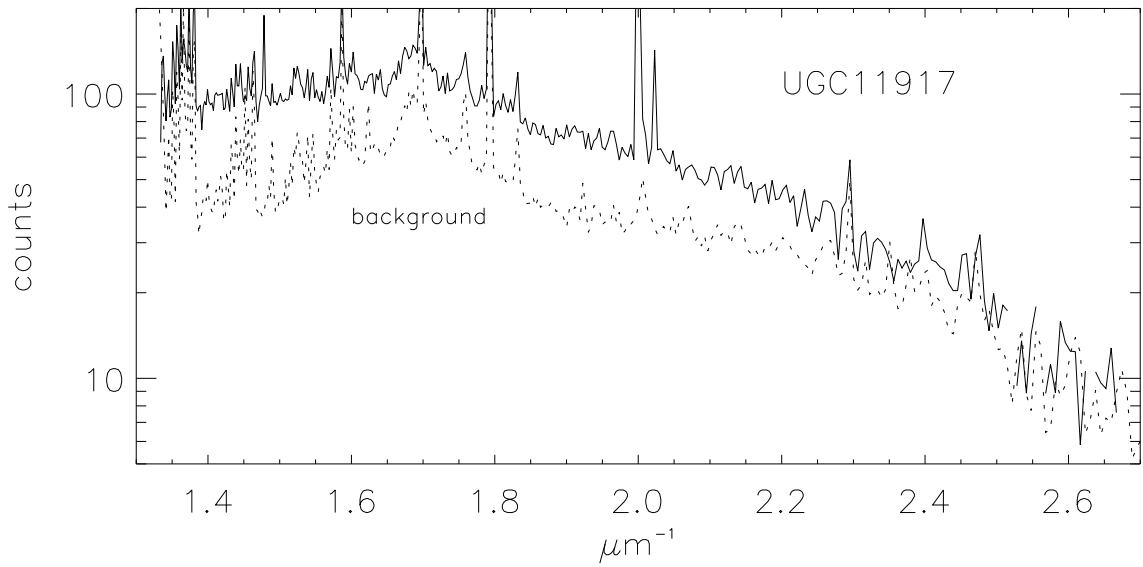}} 
\caption{Spectrum s69 of the 2-D array of galaxy UGC11917, 
and the background.
} 
\label{fig:gal}
\end{figure*}

The spectrum of galaxy UGC11917 (Figure~\ref{fig:gal}) 
is representative of many spectra of 
faint extended galaxies observed with FAST (see the FAST data-base 
at: http://tdc-www.harvard.edu/archive/), 
which all have a similar cut in intensity around $1.7\,\rm \mu m^{-1}$.¥ 
The light from UGC11917 is faint, just above the background.
It spreads $\sim 20''$ (the UGC catalogue 
gives a minor axis of UGC11917 of $24''$ in the red) on the sky.
There are three identical spectra, $s68$, $s69$, $s70$, on each side 
of which the signal decreases to the background. 
Longward of $\sim 2.2\rm \mu m^{-1}$ (shortward of $4500\,\rm \AA$), 
the galaxy spectra meet the background, meaning that, 
in the near-UV, the galaxy is nearly totally extinguished by the atmosphere.

The slit in NGC6309 and NGC6891 observations contains or is close to the central 
star (Table~\ref{tbl:kittpeak});
both 2-D arrays will most certainly 
include spectra dominated by light from the central star. 

\begin{figure*}[p]
\resizebox{1.\textwidth}{!}{\includegraphics{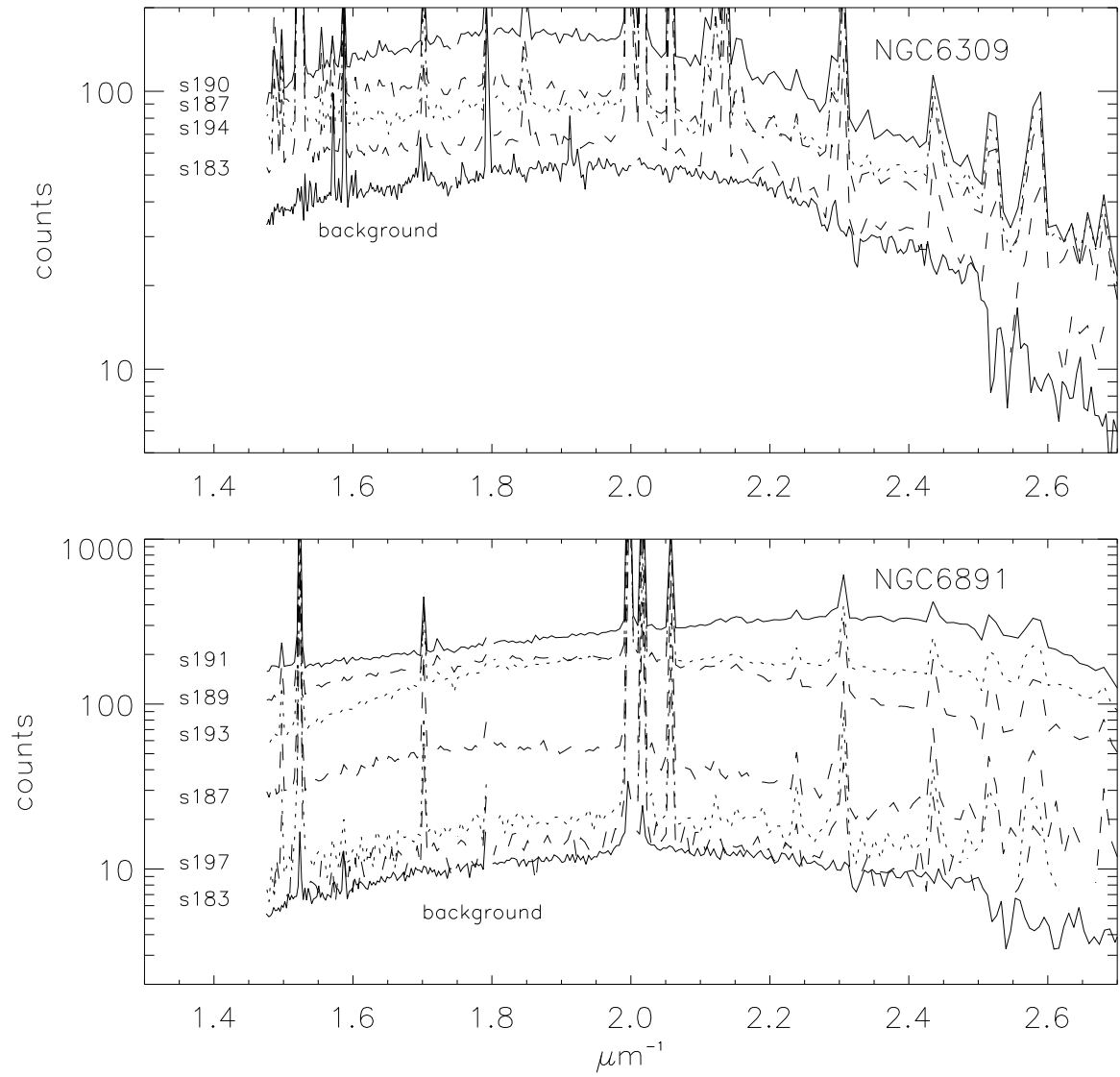}} 
\caption{Spectra extracted from the 2-D arrays of NGC6309 and NGC6891.
} 
\label{fig:neb1}
\end{figure*}

\begin{figure*}[]
\resizebox{\textwidth}{!}{\includegraphics{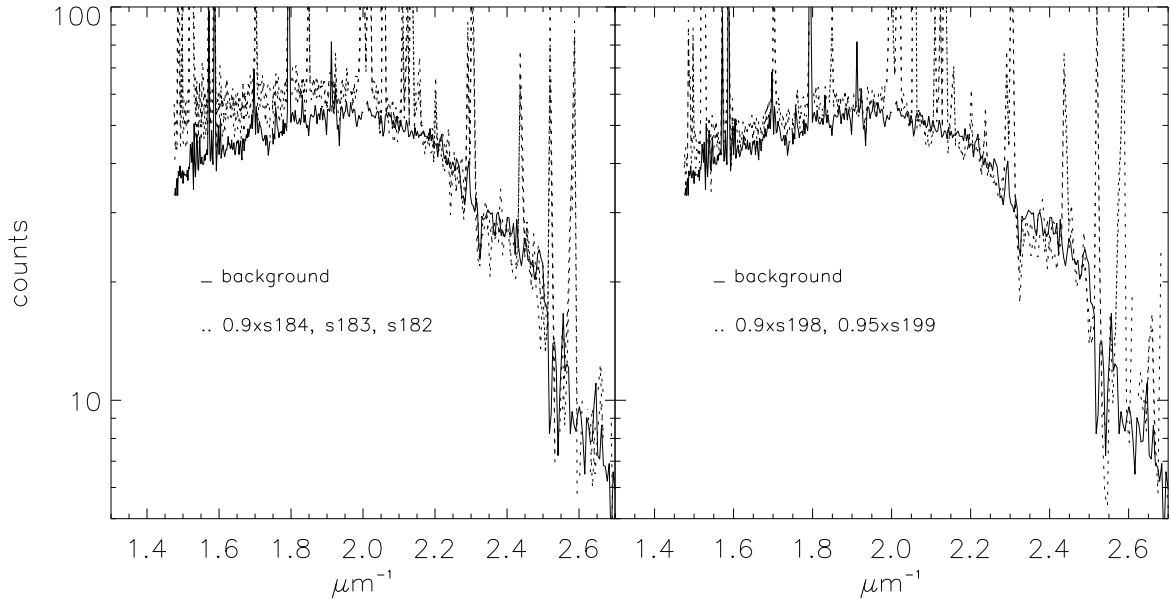}} 
\caption{Relation between the background and spectra $s184$, $s183$, 
$s182$ (left plot), $s198$, $s199$ (right plot), 
from NGC6309 observation.
An offset of 5~counts was subtracted to all spectra. 
This offset, with negligible effect in the red, is 
necessary for the spectra to match the background for 
$1/\lambda >2.5\,\rm\mu m^{-1}$.
It may correspond to blue light from the nebula since it has to be 
subtracted on a spatial extent which agrees with the size of 
NGC6309.
} 
\label{fig:ngc6309}
\end{figure*}

The main spectrum for the 2-D array of NGC6309 is $s190$ 
(Figure~\ref{fig:neb1}).
The shape of the spectrum undergoes strong variations over 
five pixels (4'') on each side of 
$s190$ ($s185$ to $s195$), 
with an increase of the blue slope with decreasing pixel 
number (as in the star observations).  
The shapes of $s184$ to $s182$ and of $s197$ to $s199$ differ from the 
background only in their reddest part ($1/\lambda <2\,\rm\mu 
m^{-1}$, Figure~\ref{fig:ngc6309}).
$s182$ and $s181$, $s200$ and $s201$ are nearly identical to the background, which 
is reached at pixels $180$ and $202$.
The signal is decreased by four between $s190$ and the 
background.

The main spectrum of NGC6891 is $s191$ (Figure~\ref{fig:neb1}).
The background, $\sim 40$ times less than $s191$,
is reached at pixels 199 and 180.
From $s191$ to $s194$,
the spectra decrease in the red more quickly than in the blue;
$s195$ to $s199$ are proportional to the 
background.
From $s191$ to smaller pixel numbers (pixels 190 and 189),
the decrease is first more pronounced in the blue.
The blue spectra ($1/\lambda>2\,\rm\mu m^{-1}$) of $s189$ to $s186$
are proportional to the background, the spectra differ in the red.
Spectra $184$ to $181$ are nearly proportional to the background, with 
a remaining, but small, excess in the red.

In the blue, the pixel range over which the shape of the spectra of NGC6891 and 
NGC6309 differs from that 
of the background (spectra $s185$ to $s194$ for NGC6891, $s185$ to $s195$ for 
NGC6309) represents $\sim 10$ pixels ($\sim 8$"), which is half 
what is expected from the angular extent of the nebulae ($\sim 15$" 
or $\sim 20$~pixels).
In the red, NGC6891 and NGC6309 differ from the 
background over $\sim 15$" (pixels $181$ to $198$ for NGC6891,
pixels $182$ to $199$ for NGC6309),
which is about the size of the nebulae.
We conclude that nebular light from these two observations is restricted to -at most- a
red excess, and a possible small offset in the blue (Figure~\ref{fig:ngc6309}).
\section{Analysis of the background in the observation of stars} 
\label{fdstar}
\begin{figure}[]
\resizebox{\columnwidth}{!}{\includegraphics{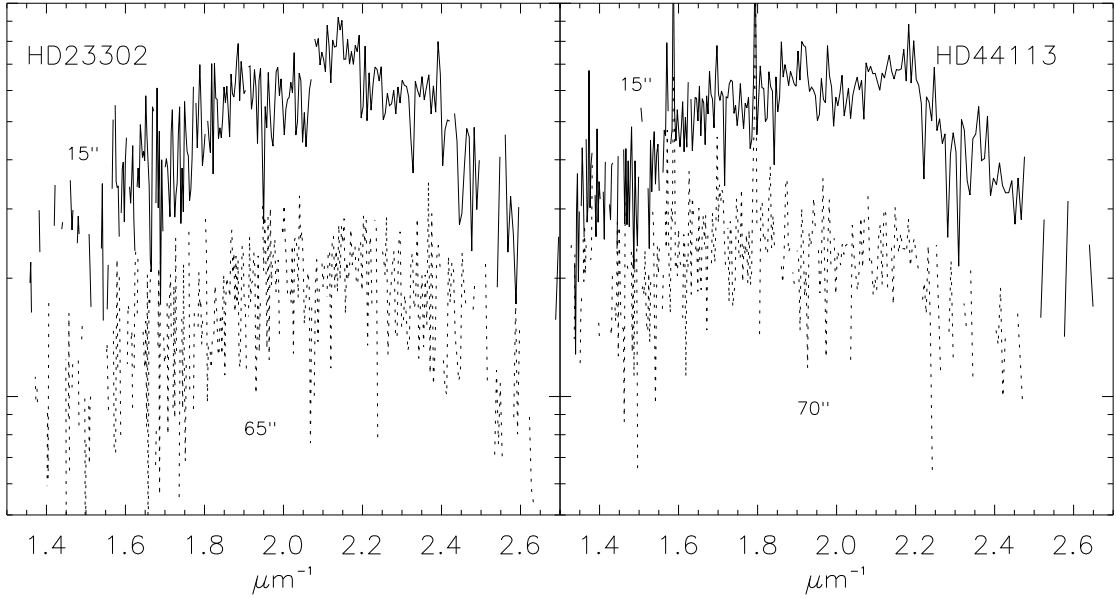}} 
\caption{Left is an average of 10 and 40 backgrounds at mean distances 
15" and 65" from HD23302.
Right is an average of 10 and 20 backgrounds at mean distances 
15" and 70" from HD44113.
The background for HD23302 decreases but keeps an identical shape 
through out the 2-D array. 
In the backgrounds from HD44113 observation, one can see the change of 
shape after the decrease in intensity.} 
\label{fig:fdec}
\end{figure}

Because of the noise level, it is difficult to make a distinction 
between individual background spectra in the observation of a star.
However, adding a sufficient number of these spectra around 10" and 65" 
from the direction of HD23302, 
a difference will be perceived which proves that the 
background, if it keeps a constant shape, diminishes with increasing 
distance from the direction of the star (Figure~\ref{fig:fdec}, left). 
Adding even more spectra at the edges of the slit for HD44113, we see 
(Figure~\ref{fig:fdec}, right)
that the shape will also ultimately change.

\begin{figure*}[t]
\resizebox{1.\textwidth}{!}{\includegraphics{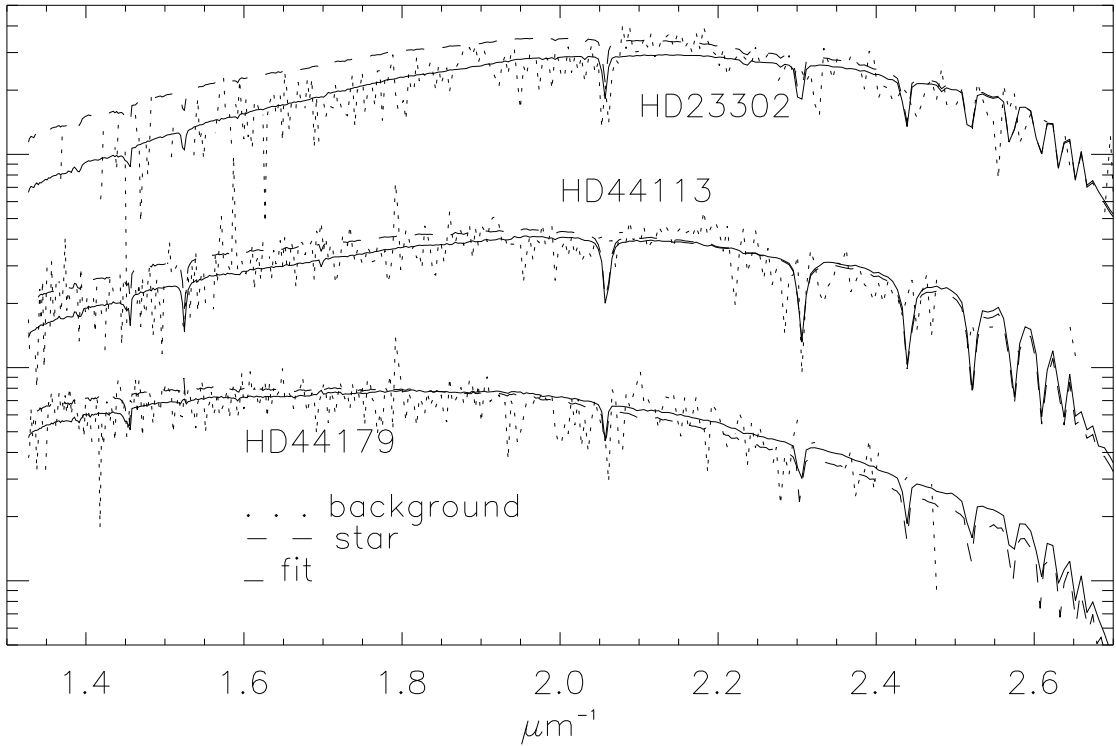}} 
\caption{Fit of the star observations backgrounds.
\emph{Dots:} background spectra
(first December 2001 observations for HD44179 and HD23302, February 
2002 observation for HD44113). 
\emph{Dashes:} 1-D spectra 
of the star (scaled by a constant factor).
\emph{Solid lines:} spectra of the stars multiplied by 
$e^{-a/\lambda}/\lambda$ (and a constant factor).
$a$ is 0.22, 0.2, 0.1 for HD44179, HD44113, HD23302.
} 
\label{fig:fdfit}
\end{figure*}

The straightforward interpretation of the continuum of the background 
spectra is that it is starlight scattered in the atmosphere, which 
explains the decrease of the background
with angular distance from the star.
Following Zagury \& Goutail (2003), since the spectra are observed close 
to the direction of the stars, the scattering is expected to be 
forward scattering by stratospheric aerosols, with an analytical 
dependence of the spectra as $1/\lambda e^{-a/\lambda}$.
Figure~\ref{fig:fdfit} gives the fit found for each background.
\section{The background in the observation of extended objects} 
\label{fdneb}
\begin{figure*}[]
\resizebox{1.\textwidth}{!}{\includegraphics{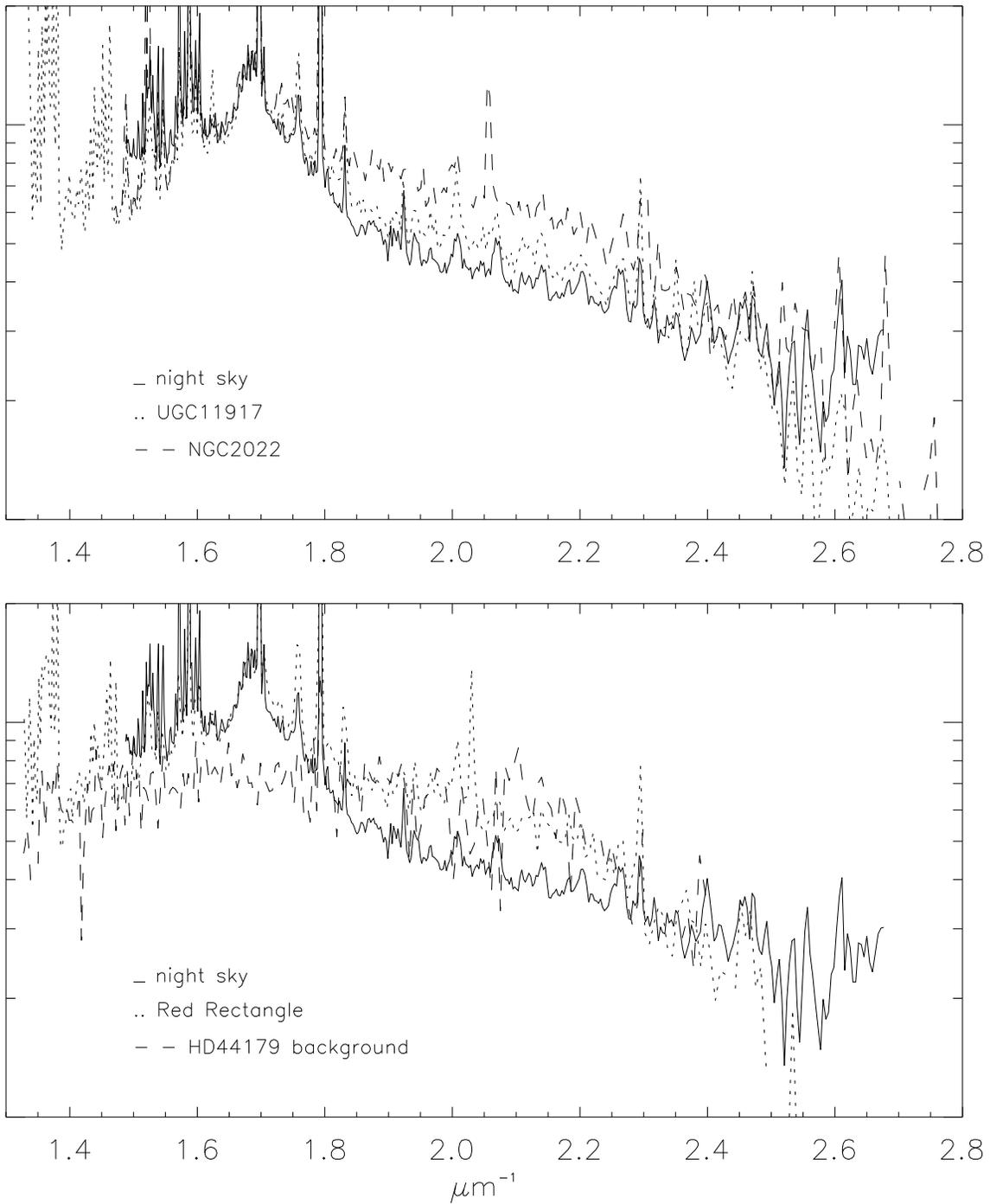}} 
\caption{Comparison of extended objects backgrounds to Massey \& 
Foltz (2000) night sky spectrum.
Upper panel shows that UGC11917's background nearly equals the night sky, while 
there is an excess in the blue for the background of NGC2022 
observation.
For the Red rectangle nebula, this excess corresponds to the background found in the observation of 
the star and can be attributed to starlight scattered in the 
atmosphere (lower panel).
} 
\label{fig:nebbg}
\end{figure*}

The backgrounds in the Red Rectangle nebula, in NGC2022, and in galaxy UGC11917
are proportional in the red, where they coincide with the night sky spectrum 
of Massey \& Foltz (2000) (Figure~\ref{fig:nebbg}).

In the blue, UGC11917 background nearly equals the night sky spectrum, while 
NGC2022 and the Red Rectangle nebula are in excess.
In the case of the Red Rectangle, comparison with observations of 
HD44179 shows that the blue excess is well fitted by HD44179 
observations background (Figure~\ref{fig:nebbg}, lower panel),
i.e. that it is attributable to light from
HD44179 scattered in the atmosphere.

We deduce that in the observations of UGC11917, NGC2022, 
and the Red Rectangle, the background has reached the night sky 
spectrum, apart, for NGC2022 and for the Red Rectangle, 
for an excess in the blue, due to light from the central 
star scattered in the atmosphere.

The observation of NGC2022 is then understood as being mainly light 
from HD37882 scattered in the atmosphere for spectrum $s205$, which 
must correspond to the closest position to HD37882 on the slit.
Moving away from pixel 205, the night sky spectrum progressively 
appears when the scattered light component diminishes.

The same diminution of the scattered light is observed between
the background spectra of the Red 
Rectangle nebula close to the nebula, and at the edges of the 
slit. 
Using the complete set of FAST observations, this will clearly be 
established 
in the following article on the Red Rectangle. 

The background in NGC6309 and NGC6891 observations is not the night 
sky, and can only be light from the central stars scattered in the atmosphere.
Spectra from these 2-D arrays are all (except for the 
small excess found in the red for some of the spectra) light from the central stars 
refracted or scattered in the atmosphere.
\section{Discussion} 
\label{dis}
The analysis conducted so far shows that the action of 
atmospheric extinction on starlight can be observed up
to more than one arcminute from the direction of a star: 
direct light from a star, refracted in the atmosphere, and scattered 
starlight by aerosols, will be felt until their level passes under that of 
the night sky.

Over a radius of a few arcseconds (3-8" for the observations presented here) 
around the direction of the star, direct 
light from the star, which has followed different paths through the atmosphere, 
dominates the spectrum.
In this angular region, the orientation of the slit in the azimuth-altitude plane 
probably determines 
the variations of the shape of the spectrum.

After this central angular region, the spectrum has decreased by a 
factor of $\sim 100$; starlight forward-scattered 
by aerosols in the atmosphere contributes most.
The shape of the spectrum, which is proportional to the spectrum of 
the star times $1/\lambda$, remains constant.
The distribution of the scattered light is spherically symmetric (it 
is the same on both sides of the slit when moving away from the star).
Its intensity decreases with distance from the star.
Starlight scattered by aerosols will ultimately fade 
under the background night sky.
The ratio of scattered starlight (which depends on 
the star magnitude) to night sky (independent of the star),
as a function of distance from the star,
should be fixed by the magnitude of the star.

If the star illuminates a nebula,  
lights scattered by the nebula and by the atmosphere will 
compete in the observed spectrum, and, at small angles, with 
direct light from the star.
Contrary to the ratio of starlight scattered in the atmosphere to 
night sky, the ratio of starlight scattered in the atmosphere to 
nebular light should not depend on the magnitude of the illuminating 
source.
It will depend on the degree of atmospheric extinction (thus on 
wavelength, on airmass, and on atmospherical conditions), since, 
if atmospheric extinction is low, light from the nebula will more 
easily reach the observer, 
while starlight scattered in the atmosphere will be diminished. 

It is infered that nebular light will more easily be perceived 
in the red, where atmospheric extinction is minimized,
rather than towards the UV, where Rayleigh atmospheric extinction ($\propto 
1/\lambda^{4}$) is important.
On the other hand, starlight scattered in the atmosphere will 
first appear in the 
blue, also because atmospheric extinction 
(and therefore atmospheric scattering by aerosols) is enhanced when moving to UV 
wavelengths.
In between these two wavelengths' domains one finds the ozone 
absorption band.

The continua from the 2-D arrays of the observations of nebulae 
examined in Section~\ref{nebspec} provide a direct application of these 
conclusions.
Nebular light is obvious for the Red Rectangle in the red, though in 
the blue the present analysis does not permit to separate the spectrum 
of the nebula from that of light from HD44179 scattered in the 
atmosphere, and from the night sky.
For NGC6309 and NGC6891 spectra, the part due to the nebula is small, restricted to 
an excess in the reddest part of the spectrum, eventually to an offset in the blue.
In these observations, the proximity of the central star, and 
its extinction by the atmosphere,
clearly determine the shape of the continuum, in and out of the 
nebula.
No nebular light was detected for NGC2022: the decrease of the spectra 
on each side of pixel 205 illustrates 
the passage from starlight scattered in the atmosphere to night sky 
spectrum.

Close to the direction of a star forward scattering by large 
grains (aerosols for the atmosphere) should prevail both for the 
light scattered by a nebula and for the light scattered by the 
atmosphere, with an analytical dependence $\propto 1/\lambda \, 
e^{-a/\lambda}$, as it was found for the backgrounds in the 
observation of stars.
The angle of scattering for starlight 
scattered in the atmosphere will always be extremely small.
This may not be the case for starlight scattered by 
nebulae, since the angle of scattering can be large 
if the nebula is close to its illuminating star.
Under these conditions, Rayleigh scattering 
by the gas could prevail in some nebulae, with a $\lambda^{-4}$ 
(rather than $\lambda^{-1}$) dependence on 
$\lambda$.
   \section{Conclusion}
This paper has considered different data-sets issued from 
long-slit, ground-based, spectroscopy at visible wavelengths, of stars, 
nebulae, and of one galaxy (although representative of faint extended galaxies 
observed with FAST).
The objects were separated into two classes: point-like objects, as 
stars, and extended objects.
The paper has surveyed different aspects of the influence on these 
observations atmospheric extinction has,
and it has lingered over the study of the background (i.e. the 
signal away from the object) spectra in these observations.

The study of the 2-D spectra of stars has shown that the direct light 
from a point-source covers a few arcseconds of the slit.
By a clear sky, stars' spectra are little affected by ozone absorption, while 
Rayleigh extinction causes the important bending observed at the 
shortest wavelengths ($1/\lambda>2\,\rm\mu m^{-1}¥$; $\lambda<5000\,\rm\AA$).
At larger distances, the spectrum is that of starlight 
scattered by aerosols in the atmosphere, with a wavelength 
dependence as $e^{-a/\lambda}/\lambda$ (and proportional to the spectrum of 
the star).
The spectrum of the scattered light will decrease slowly, and keep a steady shape, 
until it becomes negligible compared to the night sky spectrum. 

In the spectra of extended objects, several 
effects, of different nature, are superposed. 
They are therefore more difficult to interpret.

The 2-D spectrum of an extended object 
results from:
\begin{itemize}
    \item  Light from the region of the extended object itself, 
    sampled by the pixel under consideration, which will more easily be 
    perceived in the red (towards the near-infrared).
    \item  Direct light from the central source (if any) refracted in the 
    atmosphere, which can be efficient up to a few arcseconds from the source.
    \item  Light from the central source scattered in the atmosphere,
    which can be present over more than one arcminute, 
    mainly in the short wavelength range.
    \item  The night sky, which contributes to the spectrum when 
    the preceding sources of light are faint enough.
\end{itemize}¥
It may also be (as for point sources) that light from the nearest regions, 
refracted by the atmosphere, contributes to what is received by a given pixel.
However, since refracted light quickly decreases with distance, this 
should have a minor effect. 

For nebulae, we noted that the relative strength of light from the nebula and starlight 
scattered in the atmosphere does not depend on the magnitude of the 
illuminating star (but depends on the degree of atmospheric extinction). 

Spectra above the background, in the observations of NGC6309 and of NGC6891,
were found to be 
mainly direct light from the central sources, and light from these 
sources scattered in the atmosphere; the continuum of nebular light in 
these observations is probably not more than a small excess in the red 
and an offset in the blue. 

The 2-D array of NGC2022 shows no nebular light and was proved to be 
light from HD37882 scattered in the atmosphere for the pixels with the 
highest signal.
This observation illustrates the passage from scattered light in the 
atmosphere to night sky spectrum
when moving away from the central star.

It is only in the observation of the Red Rectangle nebula that 
nebular light clearly appears in the red.
However the contribution of light from HD44179 scattered in the 
atmosphere was not separated in this observation.
Additional observations will help, in a 
companion paper, to  precise the relative contribution of each.

It can already be seen that the determination of nebular spectra from 
ground-based observations can not be reduced to the traditional subtraction of a 
background followed by calibration by a standard star. 
In particular, the continuum of ground-based spectra of nebulae presented 
in the Kwitter-Henry database are not reliable;
this may not necessarily affect emission lines 
(only the level of the continuum is changed), studied by K. 
Kwitter and collaborators, but should not be neglected. 

The data in hand have not permitted to deepen some problems 
in the data reduction process and on the understanding of the 
background.
The main spectra of the 2-D arrays in stars observations
are related in a way which 
would merit attention and a better understanding.
The study of the background in long-slit spectroscopy initiated here
can be improved by a more precise evaluation of its dependence with 
stars' angular distance and magnitude.
\section*{Acknowledgments}
I thank R. Cutri and J. Huchra for their help in obtaining 
the FAST data. 
I am most of all indebted to Lucas Macri who
repeatedly observed the targets used in the paper.
Susan Tokarz has kindly reduced the data, and answered many 
questions on the procedures she uses, as well as on the FAST instrument.
I also thank K. Kwitter for agreeing to the release of
some of her Kitt Peak observations.
{}
\clearpage
\begin{table}[]
\caption[]{Coordinates, visual magnitudes, size of objects used in 
the paper}		
       \[
    \begin{tabular}{ccccc}
\hline
Object & $\alpha_{2000}$ & $\delta_{2000}$ & $m_{V}$ & size$^{(1)}$ \\
\hline
HD23302 & 03 44 52 & +24 06 48 & 3.7&
\\ 
HD44113 & 06 19 37 & -11 03 08 & $\sim 9$&
\\ 
HD44179 & 06 19 58 & -10 38 15 & 9.0 & 26$^{(2)}$
\\ 
NGC2022 & 05 42 06 & +09 05 10 & 10.1 & 19$^{(3)}$
\\ 
NGC6309 &17 14 04 & -12 54 36 & 11.6&15.5$^{(3)}$
\\ 
NGC6891 &20 15 09 & +12 42 17 & 12.4&15$^{(3)}$
\\ 
UGC11917 & 22 08 08 & +45 34 15 &$\sim 17$&24$^{(4)}$
\\ 
\hline
\end{tabular} 
 \]
\begin{list}{}{}
\item[$(1)$] size of extended objects ('').
\item[$(2)$] size of the Red Rectangle nebula $11$'' from HD44179.
\item[$(3)$] size of nebula, from Acker \emph{et al.} (1992).
\item[$(4)$] Minor axis, in the red, from the Uppsala Galaxy 
Catalogue (UGC).
\end{list}
\label{tbl:pos}
\end{table}
\begin{table*}[]
\caption[]{Mount Hopkins FAST observations}		
       \[
    \begin{tabular}{ccccc}
\hline
Position &  U.T.$^{(1)}$& $\Delta t^{(2)}$  & 
A.M.$^{(3)}$ & Alt.$^{(4)}$\\ 
\hline
\multicolumn{5}{c}{December 2001 (2001-12-21)}\\
HD23302-1 & 06:08:49 &  0.2  & 1.04 &  
87.18 \\ 
HD23302-2 &06:13:10 &  0.2  & 1.04 &  
87.32 \\ 
\hline
\multicolumn{5}{c}{December 2001 (2001-12-22)}\\
HD44179-1 &08:44:21 &  15  & 1.41 &  
44.92 \\ 
R.R. nebula $^{(5)}$ & 09:03:31 &  120 &  1.46 &  42.96  
 \\ 
HD44179-2 & 09:10:04 &  15 &  1.48 &  
42.44 \\ 
\hline
\multicolumn{5}{c}{February 2002 (2002-02-09)}\\
HD44113 & 5:05:06 &  15  &  1.38 &  46.28 \\ 
HD44179 & 5:07:24 & 5 &  1.37 &  46.59 \\ 
\hline
\multicolumn{5}{c}{March 2003 (2003-03-26)}\\
HD23302 & 03:02:00 &  0.2  & 1.71 &  35.84 \\ 
HD44179-1 & 3:10:46 &  5 &  1.51 &  41.40 \\ 
HD44179-2 & 3:33:27 &  5 &  1.60 &  38.52 \\ 
\hline
\multicolumn{5}{c}{October 2002 (2002-10-31)}\\
UGC11917 &04:59:03 &  1200  & 1.17 &  58.89 \\ 
\hline
\end{tabular} 
 \]
\begin{list}{}{}
\item[$(1)$] UT HH:MM:SS at start of exposure.
\item[$(2)$] Duration of exposure (seconds).
\item[$(3)$] Air mass.
\item[$(4)$] Altitude ($^{\circ}$)¥ of the object at time of observation.
\item[$(5)$] Observed 11.5' north from HD44179.
\end{list}
\label{tbl:fast}
\end{table*}
\begin{table}[]
\caption[]{Kitt Peak observations}		
       \[
    \begin{tabular}{ccccccc}
\hline
Object & Date$^{(1)}$& U.T.$^{(1)}$& $\Delta t^{(2)}$ & $\Delta \delta ^{(3)}$ & 
A.M.$^{(4)}$ & Alt.$^{(5)}$\\ 
\hline
NGC2022 & 1996-12-08 & 6:52:31 &  300 & $-6.5^{(6)}$ & 1.12 &  
62.41
\\ 
NGC6309 & 1999-07-01 & 04:21:08 &  300 & $\sim +3$ & 1.59 &  38.67  
 \\ 
NGC6891 & 1999-07-01 & 09:34:58&  30 & $\sim 1$ & 1.07 &  69.60 
\\ 
\hline
\end{tabular} 
 \]
\begin{list}{}{}
\item[$(1)$] UT date (YYYY-MM-DD) and time (HH:MM:SS) at start of exposure.
\item[$(2)$] Duration of exposure (seconds).
\item[$(3)$] Declination offset from the central star ('').
\item[$(4)$] Air mass.
\item[$(5)$] Altitude ($^{\circ}$)¥ of the object at time of observation.
\item[$(6)$] Requested offset was 6.5"~S. The header, which 
normally indicates the central position of the slit, gives 16"~S.
\end{list}
\label{tbl:kittpeak}
\end{table}
\clearpage
\clearpage
\end{document}